\documentclass[%
 aip,
 amsmath,amssymb,
reprint,%
]{revtex4-1}

\usepackage{graphicx}% Include figure files
\usepackage{dcolumn}% Align table columns on decimal point
\usepackage{bm}% bold math
\usepackage[utf8]{inputenc}
\usepackage[T1]{fontenc}
\usepackage{mathptmx}
\usepackage{etoolbox}
\usepackage{textgreek}
\usepackage{todonotes}

\makeatletter
\def\@email#1#2{%
 \endgroup
 \patchcmd{\titleblock@produce}
  {\frontmatter@RRAPformat}
  {\frontmatter@RRAPformat{\produce@RRAP{*#1\href{mailto:#2}{#2}}}\frontmatter@RRAPformat}
  {}{}
}%
\makeatother

\usepackage[normalem]{ulem}

\begin{document}

\preprint{AIP/123-QED}

\title{Barrier--channel intermixing and 2-dimensional electron gas degradation in Al-rich Al(Ga)N/AlGaN high electron mobility transistor heterostructures}
\author{P. Pampili}
\email{pietro.pampili@tyndall.ie}
\affiliation{Tyndall National Institute, University College Cork, Cork T12 R5CP, Ireland}

\author{V. Z. Zubialevich}
\affiliation{Tyndall National Institute, University College Cork, Cork T12 R5CP, Ireland}

\author{B. Mondal}
\affiliation{Tyndall National Institute, University College Cork, Cork T12 R5CP, Ireland}

\author{J. Mukherjee}
\affiliation{James Watt School of Engineering, University of Glasgow, Glasgow, G12 8LT, United Kingdom}

\author{S. Schulz}
\affiliation{Tyndall National Institute, University College Cork, Cork T12 R5CP, Ireland}
\affiliation{School of Physics, University College Cork, Cork T12 YN60, Ireland}

\author{D. A. J. Moran}
\affiliation{James Watt School of Engineering, University of Glasgow, Glasgow, G12 8LT, United Kingdom}

\author{P. J. Parbrook}
\affiliation{Tyndall National Institute, University College Cork, Cork T12 R5CP, Ireland}
\affiliation{School of Engineering, University College Cork, Western Road, Cork, Ireland}

\date{\today}% It is always \today, today,

\begin{abstract}
In this work, we report on recent results in understanding and addressing the issue of interface smearing in high-aluminium content AlGaN/AlGaN heterostructures. 
On the one hand, the growth of high-crystal quality AlGaN by metal-organic vapour phase epitaxy (MOVPE) requires the use of high temperatures, but on the other hand this may lead to alloy intermixing between barrier and channel layers, which smoothens out the polarization contrast and severely degrades or even completely destroys the 2-dimensional electron gas (2DEG).
We show that X-Ray Diffraction (XRD) analysis can be used as a non-destructive way to assess the sharpness of the interface, and that improved growth schemes can be successfully used to achieve high-quality 2DEG, as confirmed by contactless resistivity measurements. 
In particular, sheet resistivities around 2,500 $\Omega/\square$ were demonstrated for AlN/Al$_{0.75}$Ga$_{0.25}$N, consistent with the best-reported values in the literature. 

\end{abstract}

\maketitle

AlGaN/GaN high electron mobility transistors (HEMTs) have emerged as one of the most successful electronic devices for high-frequency and high-power-density applications, owing to the exceptional mobility and high carrier density of the characteristic 2-dimensional electron gas (2DEG) formed at the AlGaN/GaN interface. 
Despite the success of these devices, it has become challenging to keep up with industry demands for increased high voltage and high power operation, as the technology is fast approaching material limits.

An approach that has attracted significant interest is the replacement of GaN with Al-rich AlGaN as the channel material, in order to exploit the advantages of ultra wide bandgap semiconductors
\cite{tsaoUltrawideBandgapSemiconductorsResearch2018, bacaAlrichAlGaNBased2020, wooWideUltrawidebandgapSemiconductors2024}.
However, such AlGaN/AlGaN HEMTs are still at an early stage of development, and several challenges remain to be addressed.
In this study, we report about a new understanding of the impact of barrier--channel intermixing on the conductivity of the 2DEG, as well as a growth and characterization strategy to overcome this issue.

The samples discussed in this study were grown by metal-organic vapour phase epitaxy (MOVPE) using an AIXTRON CCS 3x2” showerhead reactor, with trimethylgallium (TMGa), trimethylaluminium (TMAl) and ammonia (NH\textsubscript{3}) as precursors for Ga, Al and N, respectively. 
Before the growth of the actual HEMT heterostructures, a number of AlN templates was grown on sapphire wafers using an approach previously reported elsewhere
\cite{liAlNHeteroepitaxySapphire2013}. 
The template thickness was set in the range 2.0–2.5 µm, which was sufficient to ensure good crystal quality, with a threading dislocation density on the order of low $10^{9}$ cm\textsuperscript{-2}, while simultaneously avoiding excessive tensile strain and suppressing crack formation, a known issue in thick AlN templates grown on sapphire
\cite{brunnerStressEvolutionAlxGa1xN2013}. 
To ensure a fair comparison, only templates with similar crystal quality were used to overgrow the HEMT structures for this study.

At the start of each run, the templates were initially heated in an ammonia atmosphere (500 sccm), and baked for 5 minutes at a temperature of 1,120 °C and pressure of 100 mbar, followed by the growth of a 200 nm AlN connecting layer, prepared under the same conditions used for the templates. 
Subsequently, a 600-nm thick AlGaN channel of different compositions was grown, followed by an Al(Ga)N barrier with higher Al content than in the channel, and a thickness of 50 nm (unless otherwise specified). 
Both barrier and channel were grown at a pressure of 50 mbar, V/III ratio of 1,000 and total gas flow of 8 slm, which resulted in a growth rate of approximately 0.58 nm/s. 
The relatively high V/III ratio was selected based on the results of a previous study
\cite{singhSurfaceMorphologyEvolution2024},
in which it led to smoother AlGaN surface morphologies compared with those obtained under standard AlGaN growth conditions, which typically use V/III ratios in the range of 50 to 100.
The growth temperature was kept at 1,160 °C for the channel, but was varied in the range 850--1,160 °C for the barrier, as will be discussed in more detail later. 
The growth temperatures here reported were measured using an \emph{in-situ} Laytec EpiCurve TT pyrometer system, which uses corrected infrared emissivity at 950 nm.

After growth, the samples were analysed for crystal quality, composition and strain-relaxation by X-ray diffraction (XRD), using a Malvern Panalytical X’Pert Pro diffractometer equipped with a 2-bounce hybrid monochromator %, selecting the Cu K-α1 wavelength,
and a PIXcell solid-state 2D detector. 
The crystal quality was estimated by the full-width at half maximum (FWHM) of the ω-scans of symmetric 002 and skew-symmetric 101 reflections, which led to peak broadening below 300 and 650 arcsec, respectively. 

For analysis purposes, in addition to the abovementioned AlN templates optimized for crystal quality, bufferless auxiliary substrates were co-loaded in almost all growth runs. 
These substrates were commercially sourced templates, in which a thin AlN nucleation layer of approximately 25 nm was sputtered on sapphire wafers. 
The HEMT structures overgrown on these bufferless substrates consistently exhibited FWHM values of the order of 2,400 arcsec for the 101 reflection of the AlGaN channel.
This is more than three times larger than those measured in HEMT structures grown on AlN templates with thick buffers, and corresponds to roughly an order of magnitude increase in edge dislocations.
The formation of a very high dislocation density in these samples was intentional, in order to promote relaxation in the channel through dislocation bending \cite{romanovStressRelaxationMismatched2003} 
and to enable separation of the AlN XRD reflections of the barrier from those of the buffer, as discussed later.

To determine composition and relaxation state of the different samples, reciprocal space maps (RSMs) of the asymmetric 105 reflection were collected.
To compensate for alignment errors that produce random ω offsets among different samples, $\bar{1}05$ reflections (i.e., the reflections with opposite ω–θ offsets) were also measured, and the RSMs were rotated by the angle necessary to align both reflections at the same Q\textsubscript{z} while maintaining opposite Q\textsubscript{x}.
Composition and degree of relaxation were calculated using the lattice constants and elastic properties reported in Ref.
\onlinecite{mondalInterplayCarrierDensity2025}
and references therein. 
Processing and analysis of the RSM data was done using our in-house developed open-access \verb|xrdpy| Python package\cite{zotero-16966}. The thicknesses of the barrier layers were measured, when possible, by X-ray reflectivity (XRR). 

In the first sample of this study, a pure AlN barrier was grown immediately after the channel at a temperature of 1,160~°C. The sample was generally smooth, with only a few cracks visible by optical Nomarski microscopy, mostly at the wafer edges, caused by a somewhat thicker AlN buffer; this issue was corrected in the subsequent samples.

\begin{figure}
    \includegraphics[width=1\columnwidth]{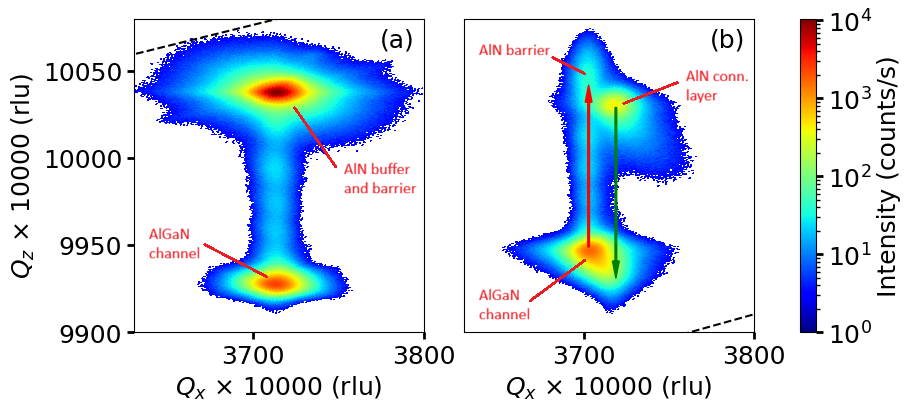}
    \caption{\label{fig:XRD_streak} Reciprocal space maps of the AlN/Al$_{0.8}$Ga$_{0.2}$N HEMT structures grown on (a) standard, and (b) auxiliary templates, with a nominal barrier thickness of 50 nm. Arrows explained in the text.}
\end{figure}

As shown in Fig.~\ref{fig:XRD_streak}(a), two main peaks are apparent in the RSMs, which correspond to AlN (barrier and buffer) and AlGaN (channel). 
The vertical alignment of the two peaks indicates that barrier and channel remained coherently strained to the buffer, and a channel composition of 80\% AlN was calculated. 
A particularly interesting feature of this RSM is the presence of a high-intensity streak connecting the two peaks, consistent with a relatively thick graded interface. 
This interpretation is further supported by the absence of Kiessig fringes in the XRR measurements
\cite{moramXrayDiffractionIIInitrides2009}.

Although sharp heterointerfaces also produce characteristic streaks called crystal truncation rods (CTRs)
\cite{robinsonCrystalTruncationRods1986}, 
their intensity typically follows a power-law decay with increasing Q\textsubscript{z} distance from the main Bragg peaks, which is not observed in our samples. Instead, the nearly constant intensity of the streak connecting the AlN and AlGaN peaks suggests the presence of a graded layer with an approximately linear variation in composition.

To rule out any effect due to the relatively high V/III ratio, the same nominal structure was grown also with the AlN barrier deposited at a V/III ratio of 50.  No detectable differences were observed compared with the sample in which the entire HEMT structure was grown at a V/III ratio of 1,000.

In order to understand if the interface grading observed in the RSM was at the buffer--channel, channel--barrier or at both interfaces, the  co-loaded sample grown on the bufferless auxiliary template was also analysed, and the results are shown in Fig.~\ref{fig:XRD_streak}(b). 
As highlighted by the two arrows, the interface between the AlN connecting layer and the channel does not show any streak (green arrow), which is consistent with a rather sharp interface. 
However, the barrier/channel interface shows the same high-intensity streak (red arrow) already observed in the strained sample.

One possible explanation for this behaviour, namely the formation of a sharp interface when the TMGa source is opened and a graded interface when it is closed, is an anomalously long residence time of the Ga precursor due to gas recirculation within the growth chamber.
To test this hypothesis, the same structure was grown again, with the introduction of a 5-minute pause between the end of the channel and the start of the barrier growth. 
The growth pause was implemented by switching off both TMGa and TMAl sources while keeping the supply of ammonia on.
Despite this growth pause being many orders of magnitude longer than the typical residence time of the species in the reactor, usually considered to be of the order of a few tenths of a second 
\cite{liMassTransportAnalysis2011}, 
no detectable improvement was found.

\begin{figure}
    \includegraphics[width=1\columnwidth]{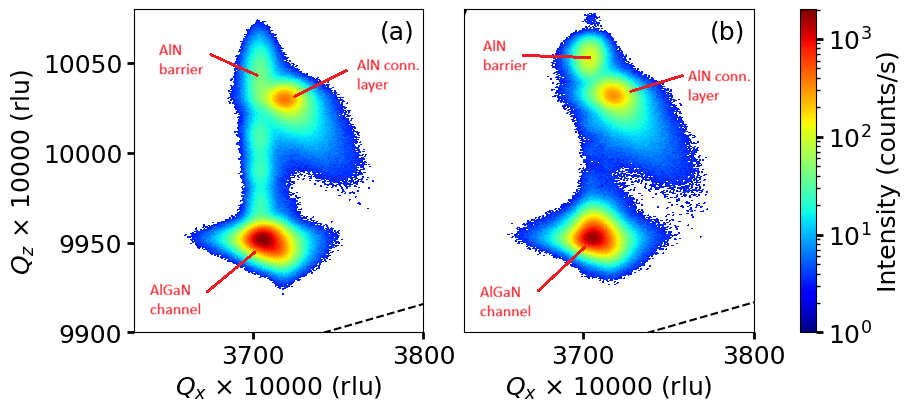}
    \caption{\label{fig:low_temp} Reciprocal space maps of the AlN/Al$_{0.8}$Ga$_{0.2}$N HEMT structures grown on the auxiliary substrate, with barriers grown at (a) 1,160 °C, and (b) 850 °C, with a nominal thickness of 50 nm. To ensure a fair comparison, sample (a) underwent a growth pause equal in duration to the time required to cool from 1,160 °C to and 850 °C during the growth of sample (b).}
\end{figure}

Although there is currently limited data on intermixing in AlGaN/AlGaN heterostructures for electronic applications, this issue is already well known in the field of UV lasers, where degradation of waveguide confinement due to smeared interfaces has been reported \cite{yoshikawaUsingLowtemperatureGrowth2023, saitoImpactGrowthTemperature2025}. 
Theoretical mechanisms have also been proposed to explain this behaviour \cite{kangawaInfluenceIntrinsicPoint2025}.
Both experimental observations and theoretical considerations suggest that low-temperature growth can be used to overcome this issue. 
For this reason, the growth was repeated using the same heterostructure, but with the barrier grown also at 1,000 °C and at 850 °C.
While the sample with a barrier temperature of 1,000 °C showed an RSM indistinguishable from the sample entirely grown at 1,160 °C, the RSM of the sample with the barrier grown at 850 °C does not present the high-intensity streak, but only much lower intensity truncation tails as shown in Fig.~\ref{fig:low_temp}.

\begin{figure}
    \includegraphics[width=1\columnwidth]{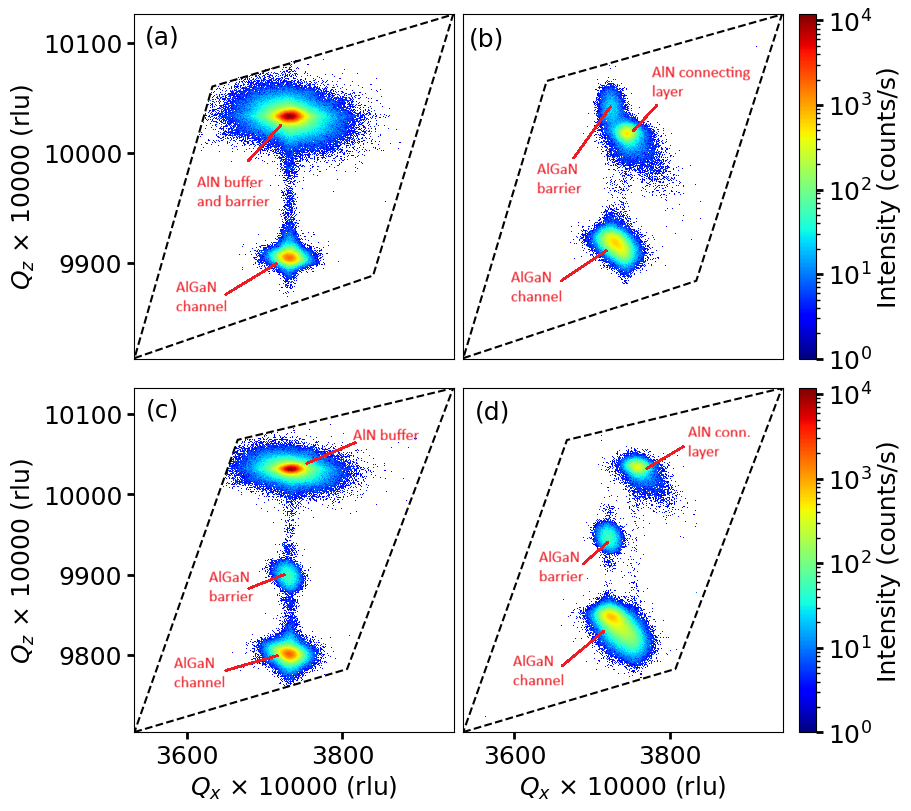}
    \caption{\label{fig:final} Reciprocal space maps of different HEMT structures grown at 850 °C, and with a nominal barrier thickness of 30 nm: AlN/Al$_{0.75}$Ga$_{0.25}$N on (a) standard, and (b) auxiliary templates; Al$_{0.75}$Ga$_{0.25}$N/Al$_{0.55}$Ga$_{0.45}$N on (c) standard, and (d) auxiliary templates.}
\end{figure}

We note that this approach also proved to be an effective solution for devices with thinner barriers and different compositions, as shown in the RSMs of Fig.~\ref{fig:final}, where high-intensity streaks are not present. Clear XRR Kiessig fringes were also observed (not shown) confirming sharp interfaces.

\begin{figure}
    \includegraphics[width=1\columnwidth]{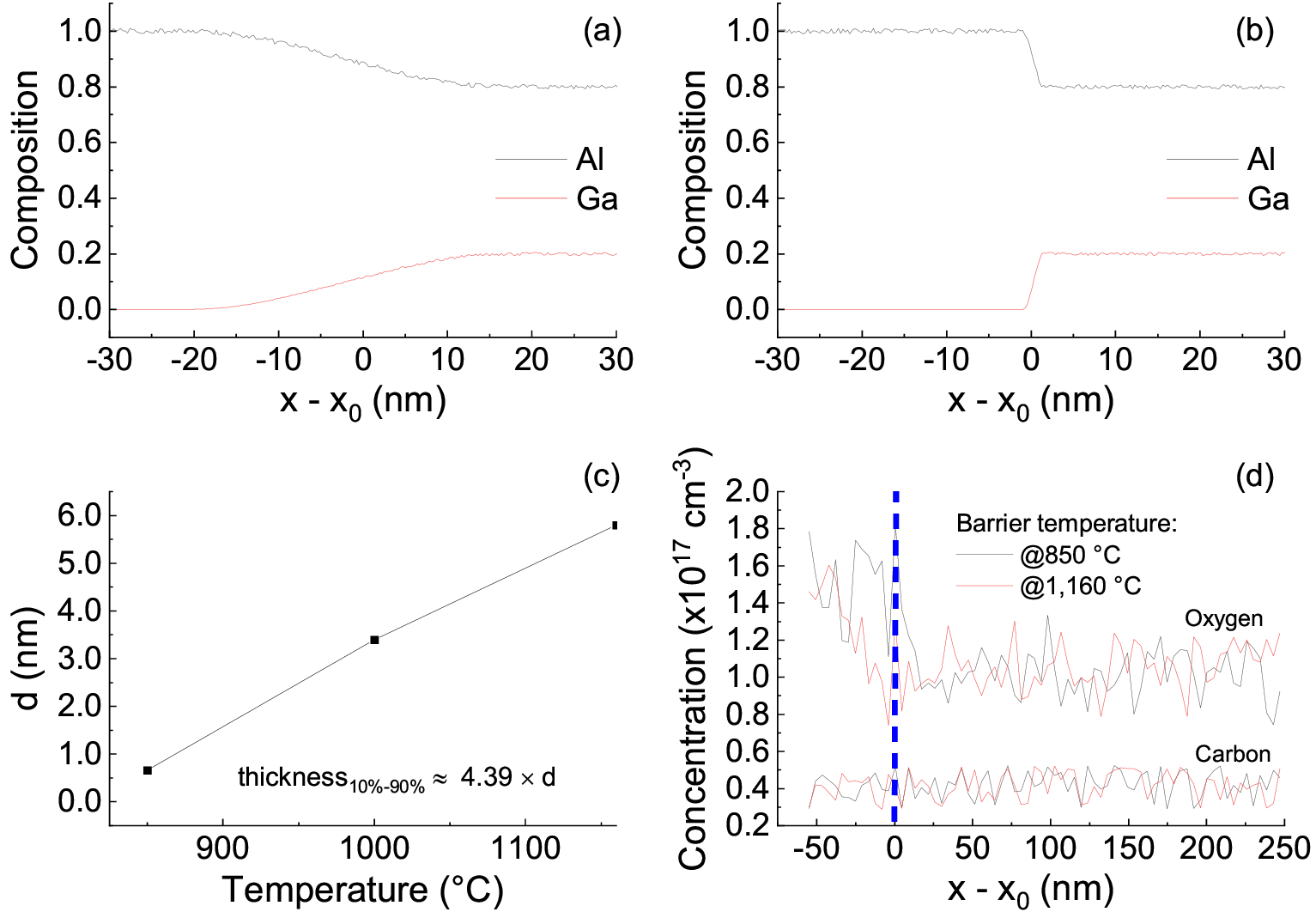}
    \caption{\label{fig:SIMS} High-resolution SIMS scans of the barrier/channel heterointerface for the AlN/Al$_{0.8}$Ga$_{0.2}$N HEMT heterostructure grown at (a) 1,160 °C, and (b) 850 °C; (c) estimation of the temperature dependence of the broadening parameter $d$ of the graded layer; (d) standard resolution SIMS scans of oxygen and carbon impurities, the vertical dashed line (blue) indicates the position of the AlN/AlGaN interface.}
\end{figure}

To further prove that the absence of a streak in the RSM is associated with a much sharper interface, we sourced high-resolution secondary ion mass spectroscopy (SIMS) measurements for samples grown on auxiliary substrates with different barrier temperatures. 
The SIMS profiles of the samples at both ends of the temperature range are shown in Fig.~\ref{fig:SIMS}(a, b).
A graded composition transition layer is apparent in both; however, while its thickness is about 35 nm in the sample entirely grown at 1,160 °C, it reduces to only about 5 nm when the barrier is grown at 850 °C. 
To better estimate the temperature dependence of the transition layer thickness, the SIMS data of all samples were fitted with a Boltzmann sigmoidal function
\begin{equation}
 y (x)= A_2 + (A_1 - A_2) \frac{1}{1+e^{\frac{x - x_0}{d}}},
\end{equation}
where $y$ is the SIMS signal for either Al or Ga compositions, as a function of the depth $x$ (centred at $x_0$), $A_1$ and $A_2$ the signal intensities of the two plateaus, and $d$ the broadening parameter, which is proportional to the graded layer thickness. 

As shown in Fig.~\ref{fig:SIMS}(c), we found that the thickness of the graded layer increases almost linearly with temperature.

Although low-temperature growth proved to be essential for achieving sharp heterointerfaces, there is concern that this may favour the incorporation of unwanted point defects. 
However, as shown by the SIMS data in Fig.~\ref{fig:SIMS}(d), the concentrations of carbon and oxygen impurities did not increase detectably with the decrease in barrier growth temperature.
Carbon levels remained constant around $4.0 \times 10^{16}$ cm\textsuperscript{-3}; oxygen levels remained constant around $1.0 \times 10^{17}$ cm\textsuperscript{-3} in the channel, and increased to about $1.8 \times 10^{17}$ cm\textsuperscript{-3} in the barrier due to the higher Al content in both samples, regardless of the temperature.
While these results suggest no dramatic increase in C and O impurities, further work is still needed to determine optimal growth conditions that limit interface smearing while minimizing point defect incorporation.

\begin{figure}
    \includegraphics[width=\columnwidth]{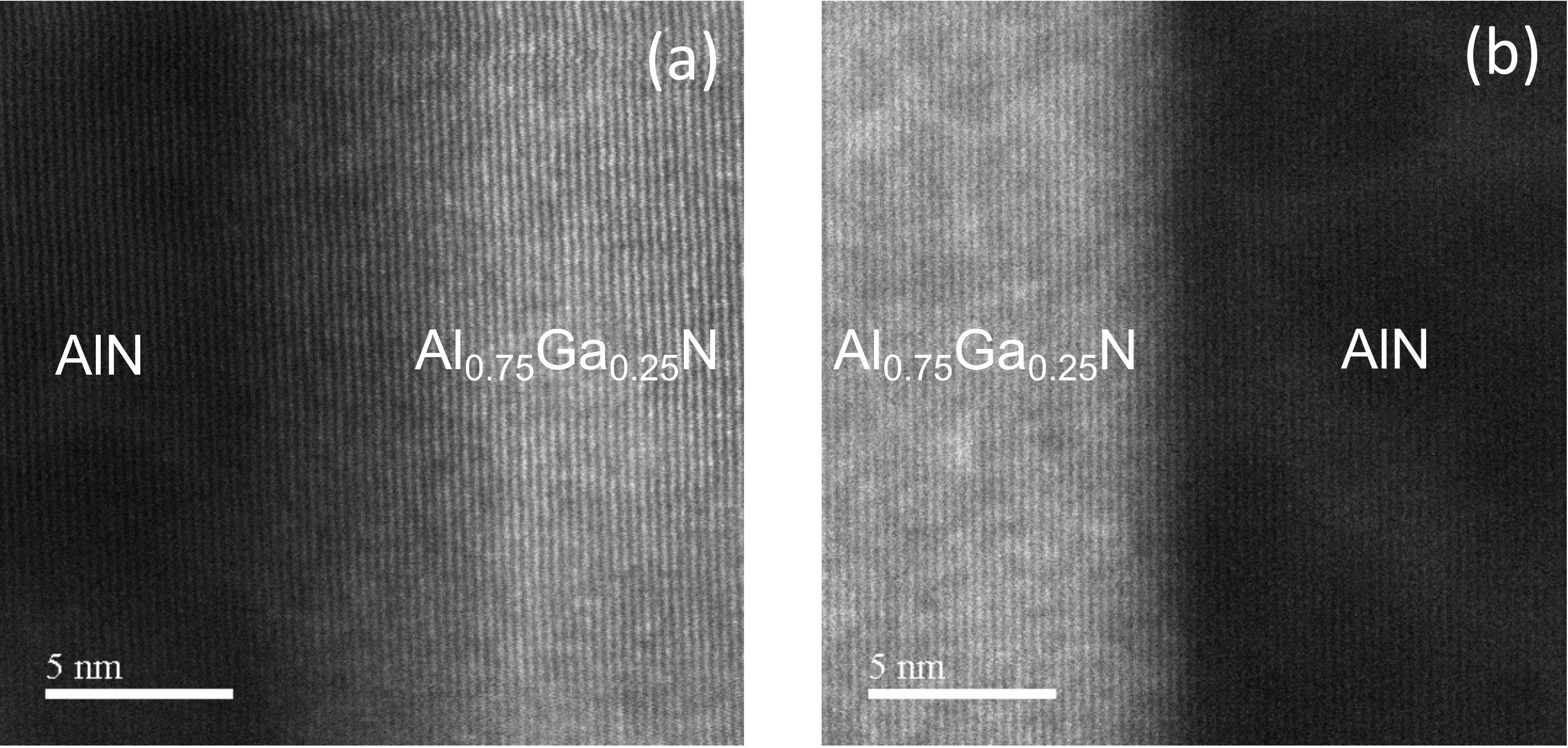}
    \caption{\label{fig:TEM} Cross-sectional TEM of the top (barrier/channel) and bottom (channel/AlN connecting layer) heterointerfaces for the AlN/Al$_{0.75}$Ga$_{0.25}$N HEMT structure: (a) top interface grown at 850~°C, and (b) bottom interface grown at 1,160~°C.}
\end{figure}

The presence of a residual ($\sim$5 nm) graded layer in the top heterointerface of the low-temperature sample is also confirmed by cross-sectional transmission electron microscopy (TEM), as reported in Fig.~\ref{fig:TEM} for the AlN/Al$_{0.75}$Ga$_{0.25}$N HEMT structure. 
It should be noted that the bottom heterointerface is much sharper despite the fact it was grown at the standard temperature of 1,160 °C, which is consistent with the lack of streak along the green arrow already mentioned in the discussion of Fig.~\ref{fig:XRD_streak}(b).

Having established the temperature-dependent origin of the composition gradient at the barrier--channel interface, we next investigated its impact on the 2DEG conductivity, a key determinant of HEMT performance. We used the Semilab LEI-88 non-contact sheet resistivity measurement system to characterize the 2DEG conductivity, which has a nominal calibration range of 16--3,600 $\Omega/\square$, but is capable of measuring up to 20 k$\Omega/\square$, although with a reduced precision characterized by a standard deviation of 17\%. 
We note that while Hall measurements provide a more accurate quantitative insight into the 2DEG properties (density and mobility), the fabrication of reliable ohmic contacts on Al-rich materials remains challenging to date.\cite{bacaAlrichAlGaNBased2020} Therefore, to gain initial insight, we report the electrical characterization based on this contactless technique. 

The contactless measurements revealed that despite the good surface morphology (RMS roughness $< 5$ nm), and low C and O densities of $\sim 10^{17} $cm$^{-3}$, as shown in Fig.~\ref{fig:SIMS}(d), none of the HEMT structures with barriers grown at 1,160 °C exhibited measurable conductivity (sheet resistivity above detection limit). However, all HEMTs with barriers grown at 850 °C showed clear 2DEG conductivity. In particular, $R_{sh}$ of about 2,500 $\Omega/\square$ and 5,500 $\Omega/\square$ were obtained for the AlN/Al$_{0.75}$Ga$_{0.25}$N and Al$_{0.75}$Ga$_{0.25}$N/Al$_{0.55}$Ga$_{0.45}$N HEMT structures, respectively. These $R_{sh}$ values are among the best reported for HEMTs with comparable composition contrasts and barrier thicknesses.\cite{bacaAlrichAlGaNBased2020}

To understand the connection between the interface composition gradient and the observed 2DEG conductivity degradation, we next performed one-dimensional Schr\"odinger-Poisson simulations. Details of the simulation methodology and parameters are provided in the Supplementary Material (Secs.~S1--S3).\cite{mondalInterplayCarrierDensity2025,Birner2007Nextnano:Simulations, Trellakis2006TheResults,Jogai2003InfluenceTransistors,*Fonoberov2003Excitonic,*Tsai1999Theoretical,*Rinke2008Consistent,*Armstrong2016Polarization-inducedAlloys,*NSMSemiconductors,*Arehart2006EffectCharacteristics,*Qiao2000DependenceFraction,*Wei1998CalculatedOrbitals,*Vurgaftman2003BandSemiconductors,*Vurgaftman2001BandAlloys,*Hahn2013AlNProposal}

\begin{figure}
    \includegraphics[width=0.98\columnwidth]{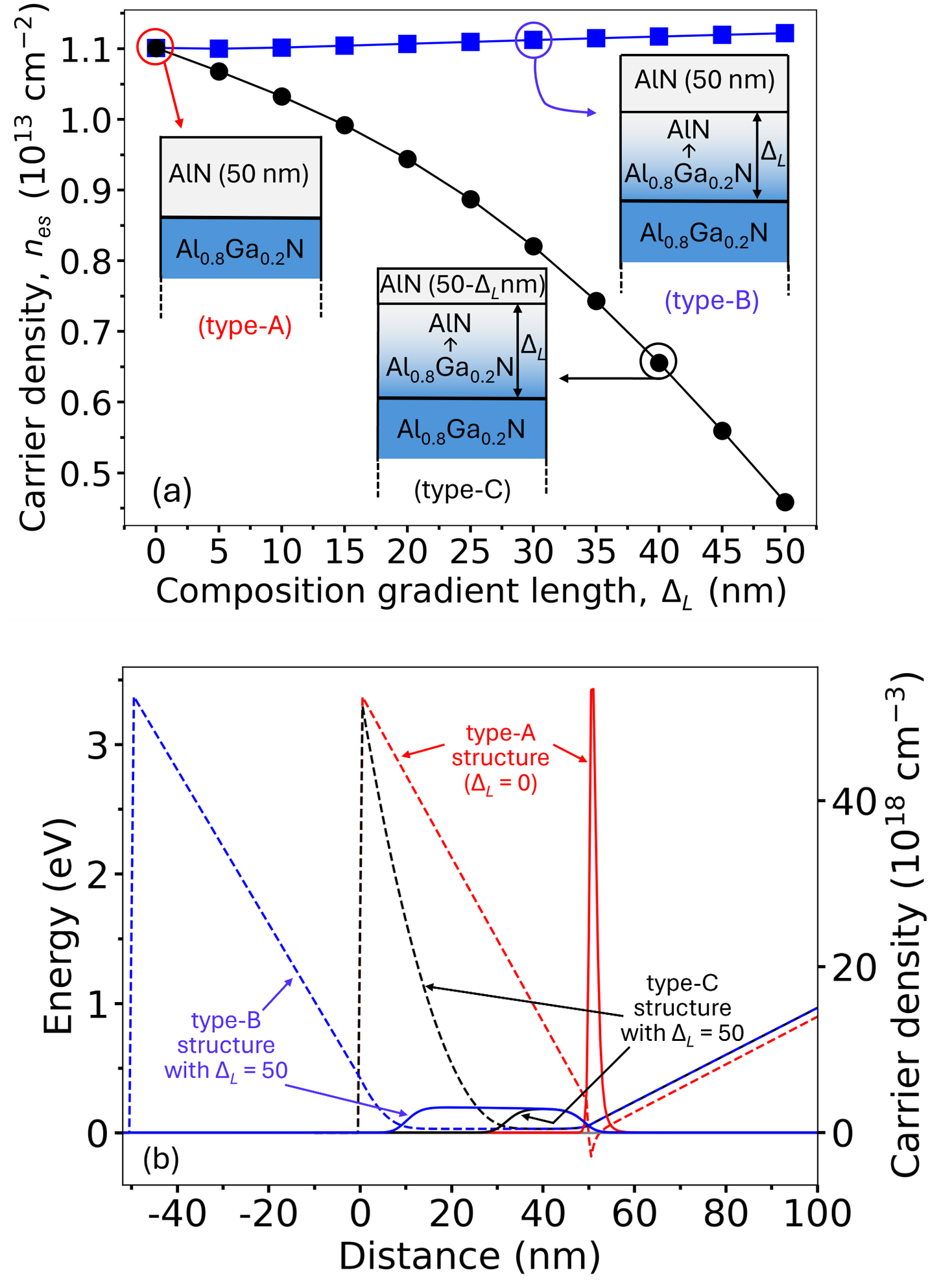}
    \caption{\label{fig:Theory} (a) Total carrier density, $n_{es}$, as a function of the composition grading length, $\Delta_L$, in an AlN(50 nm)/Al$_{0.8}$Ga$_{0.2}$N HEMT. AlN$\leftarrow$Al$_{0.8}$Ga$_{0.2}$N denotes linear grading of the Al-composition from Al$_{0.8}$Ga$_{0.2}$N to AlN; $\Delta_L = 0$ corresponds to a sharp barrier-channel interface. (b) Conduction band (dashed lines) and volumetric carrier density profiles (solid lines) for three representative structures (type-A, type-B, and type-C) with different $\Delta_L$ and barrier thicknesses from (a).}
\end{figure}

Figure~\ref{fig:Theory}(a) shows the variation of total carrier density, $n_{es}$, as a function of the composition gradient length, $\Delta_L$, introduced at the barrier-channel interface for a representative AlN(50 nm)/Al$_{0.8}$Ga$_{0.2}$N HEMT structure. Following our experimental observations in Figs.~\ref{fig:SIMS},\ref{fig:TEM}, a \textit{linear} composition grading was incorporated \textit{within} the AlN barrier in the simulated structures, see Fig.~\ref{fig:Theory}(a) bottom-right schematic. In this case, $n_{es}$ decreased with increasing $\Delta_L$ (black circles), as the composition grading within the barrier reduces the `effective' barrier thickness.

From the carrier density distribution profiles, Fig.~\ref{fig:Theory}(b), we further found that the composition grading caused the carrier density to spatially spread out. This can affect the carrier transport: as the carrier density extends toward the binary AlN barrier within the graded region, alloy disorder related effects are reduced. An increase in the overall carrier mobility, $\mu$, is thus expected with increasing $\Delta_L$.\cite{Armstrong2016,Armstrong2019,Jena2002} Although a quantitative evaluation of $\mu$ is beyond the scope of this work, prior studies on comparable systems, such as in polarization field-effect transistors (in which composition gradients are introduced intentionally), have reported mobility enhancements up to $\sim 1.5\times$ at maximum.\cite{Armstrong2019,Armstrong2016}

With increasing $\Delta_L$,  $n_{es}$ in our case decreased by a factor of $\sim 2.5$, Fig.~\ref{fig:Theory}(a), whereas $\mu$ is expected to increase by only a factor of $\lesssim 1.5$. We thus conclude that conductivity degradation, or equivalently the increase in $R_{sh}$ ($\propto 1/(n_{es}\mu)$), in an unintentional compositionally graded HEMT can strongly be driven by the reduction in $n_{es}$. While this is consistent with trends observed in our experiments, it cannot quantitatively explain the findings. Further studies, including the mobility measurements and calculations, are required to fully elucidate the origin of the observed collapse of the conductivity.

Additionally, our simulation showed that employing a thick AlN barrier can mitigate the adverse effect of unintentional composition grading on $n_{es}$ [blue squares in Fig.~\ref{fig:Theory}(a)]. In these simulations, the composition grading was introduced as an additional transition layer with the AlN barrier thickness kept fixed at 50 nm [Fig.~\ref{fig:Theory}(a), top-right schematic]. Here, we found that $n_{es}$ remained nearly constant with $\Delta_L$. The graded region only modified the carrier distribution, transitioning it from a 2- to a 3-dimensional profile [Fig.~\ref{fig:Theory}(b)], without affecting the integrated free carrier density. This suggests that by extending the growth of the barrier, in principle, the reduction of the total carrier density can be prevented. However, since thick barriers would again pose challenges to realizing reliable ohmic contacts, this approach may not be optimal from a device-implementation perspective.

In conclusion, we observed that standard growth of Al(Ga)N barriers at high temperatures leads to smeared barrier--channel interfaces and 2DEG collapse. Low-temperature growth at 850 °C prevents the intermixing and restores 2DEG. 
With the implementation of this growth strategy, we successfully demonstrated AlN/Al$_{0.75}$Ga$_{0.25}$N and  Al$_{0.75}$Ga$_{0.25}$N/Al$_{0.55}$Ga$_{0.45}$N HEMT structures with 2DEG sheet resistivity of about 2,500 $\Omega/\square$ and 5,500 $\Omega/\square$, respectively. While the results were confirmed by both high-resolution SIMS and cross-sectional TEM, we demonstrated that the presence of an intermixed layer can also be detected with non-destructive XRD analysis. In particular, the presence of high-intensity streaks in RSMs correlates with interface grading. 
We showed that unintentional composition grading at the barrier--channel interface is detrimental to HEMT devices and should thus be avoided. Our proposed low-temperature barrier growth strategy therefore proved to be a critical approach for realizing high-quality, high-performance Al-rich \mbox{AlGaN/AlGaN} HEMTs.
\section*{Supplementary Material}
A detailed description of the simulation methodology is provided in the Supplementary Material (Sections S1-S3).
\begin{acknowledgments}
We acknowledge support of Dr Paweł Michałowski from Łukasiewicz -- Institute of Microelectronics and Photonics in Warsaw, Poland, for the high-resolution SIMS; and Dr Donald MacLaren from the School of Physics, University of Glasgow, UK, for the cross-sectional TEM.
This research is jointly supported by Taighde Éireann - Research Ireland, formerly Science Foundation Ireland (SFI) under Grant 
\mbox{No. 23/EPSRC/3888}, and by EPSRC under Grant \mbox{No. EP/X036901/1}. 
Part of this research was also supported by Research Ireland under the Grants 
\mbox{No. 21/FFP-A/9014} and \mbox{No. 12/RC/2276\_P2} .
Open access publication of this article was supported by the IReL consortium.
\end{acknowledgments}

\section*{Data Availability Statement}

The data that support the findings of this study will be made openly available in Zenodo, at https://doi.org/10.5281/zenodo.20042244.

Further data can be made available upon reasonable request from the corresponding author.

\section{References}
%\nocite{*}
\bibliography{bibliography}% Produces the bibliography via BibTeX.

@article{Interplay_Mondal_AED2025,
author = {Mondal, Badal and Pampili, Pietro and Mukherjee, Jayjit and Moran, David and Parbrook, Peter James and Schulz, Stefan},
title = {Interplay of carrier density and mobility in Al-rich (Al,Ga)N-channel HEMTs: Impact on high-power device performance potential},
journal = {APL Electronic Devices},
volume = {1},
number = {},
pages = {026117},
year = {2025},
doi = {https://doi.org/10.1063/5.0277051},
url = {https://doi.org/10.1063/5.0277051}
}

@article{Hahn2013AlNProposal,
    title = {{AlN barrier HFETs with AlGaN channels to shift the threshold voltage to higher positive values: a proposal}},
    year = {2013},
    journal = {Semiconductor Science and Technology},
    author = {Hahn, Herwig and Reuters, Ben and Kalisch, Holger and Vescan, Andrei},
    number = {7},
    month = {7},
    pages = {074017},
    volume = {28},
    publisher = {IOP Publishing},
    url = {https://iopscience.iop.org/article/10.1088/0268-1242/28/7/074017},
    doi = {10.1088/0268-1242/28/7/074017},
    issn = {0268-1242}
}

@article{Vurgaftman2001BandAlloys,
    title = {{Band parameters for III–V compound semiconductors and their alloys}},
    year = {2001},
    journal = {Journal of Applied Physics},
    author = {Vurgaftman, I. and Meyer, J. R. and Ram-Mohan, L. R.},
    number = {11},
    month = {6},
    pages = {5815--5875},
    volume = {89},
    publisher = {American Institute of Physics Inc.},
    url = {https://pubs.aip.org/jap/article/89/11/5815/488612/Band-parameters-for-III-V-compound-semiconductors},
    doi = {10.1063/1.1368156},
    issn = {0021-8979}
}

@article{Vurgaftman2003BandSemiconductors,
    title = {{Band parameters for nitrogen-containing semiconductors}},
    year = {2003},
    journal = {Journal of Applied Physics},
    author = {Vurgaftman, I. and Meyer, J. R.},
    number = {6},
    month = {9},
    pages = {3675--3696},
    volume = {94},
    publisher = {AIP Publishing},
    url = {https://pubs.aip.org/jap/article/94/6/3675/292813/Band-parameters-for-nitrogen-containing},
    doi = {10.1063/1.1600519},
    issn = {0021-8979}
}

@article{Wei1998CalculatedOrbitals,
    title = {{Calculated natural band offsets of all II–VI and III–V semiconductors: Chemical trends and the role of cation d orbitals}},
    year = {1998},
    journal = {Applied Physics Letters},
    author = {Wei, Su-Huai and Zunger, Alex},
    number = {16},
    month = {4},
    pages = {2011--2013},
    volume = {72},
    url = {https://pubs.aip.org/apl/article/72/16/2011/68146/Calculated-natural-band-offsets-of-all-II-VI-and},
    doi = {10.1063/1.121249},
    issn = {0003-6951}
}

@article{Qiao2000DependenceFraction,
    title = {{Dependence of Ni/AlGaN Schottky barrier height on Al mole fraction}},
    year = {2000},
    journal = {Journal of Applied Physics},
    author = {Qiao, D. and Yu, L. S. and Lau, S. S. and Redwing, J. M. and Lin, J. Y. and Jiang, H. X.},
    number = {2},
    month = {1},
    pages = {801--804},
    volume = {87},
    publisher = {American Institute of Physics Inc.},
    url = {https://pubs.aip.org/jap/article/87/2/801/178350/Dependence-of-Ni-AlGaN-Schottky-barrier-height-on},
    doi = {10.1063/1.371944},
    issn = {0021-8979}
}

@article{Arehart2006EffectCharacteristics,
    title = {{Effect of threading dislocation density on Ni/n-GaN Schottky diode I-V characteristics}},
    year = {2006},
    journal = {Journal of Applied Physics},
    author = {Arehart, A. R. and Moran, B. and Speck, J. S. and Mishra, U. K. and DenBaars, S. P. and Ringel, S. A.},
    number = {2},
    month = {7},
    pages = {023709},
    volume = {100},
    publisher = {AIP Publishing},
    url = {https://pubs.aip.org/jap/article/100/2/023709/145275/Effect-of-threading-dislocation-density-on-Ni-n},
    doi = {10.1063/1.2219985},
    issn = {0021-8979}
}

@misc{NSMSemiconductors,
    title = {{NSM Archive - Physical Properties of Semiconductors}},
    year = {},
    booktitle = {https://www.ioffe.ru/SVA/NSM/Semicond/index.html},
    url = {http://www.ioffe.ru/SVA/NSM/Semicond/index.html},
    note = {Last accessed 10/04/2026}
}

@article{Armstrong2016Polarization-inducedAlloys,
    title = {{Polarization-induced electrical conductivity in ultra-wide band gap AlGaN alloys}},
    year = {2016},
    journal = {Applied Physics Letters},
    author = {Armstrong, Andrew M. and Allerman, Andrew A.},
    number = {22},
    month = {11},
    pages = {222101},
    volume = {109},
    publisher = {American Institute of Physics Inc.},
    url = {https://pubs.aip.org/apl/article/109/22/222101/31942/Polarization-induced-electrical-conductivity-in},
    doi = {10.1063/1.4969062},
    issn = {0003-6951}
}

@article{Rinke2008Consistent,
  title = {{Consistent set of band parameters for the group-III nitrides AlN, GaN, and InN}},
  author = {Rinke, Patrick and Winkelnkemper, M. and Qteish, A. and Bimberg, D. and Neugebauer, J. and Scheffler, M.},
  journal = {Phys. Rev. B},
  volume = {77},
  issue = {7},
  pages = {075202},
  numpages = {15},
  year = {2008},
  month = {Feb},
  publisher = {American Physical Society},
  doi = {10.1103/PhysRevB.77.075202},
  url = {https://link.aps.org/doi/10.1103/PhysRevB.77.075202}
}

@article{Armstrong2016,
   author = {Andrew M. Armstrong and Andrew A. Allerman},
   doi = {10.1063/1.4969062},
   issn = {0003-6951},
   issue = {22},
   journal = {Applied Physics Letters},
   month = {11},
   pages = {222101},
   publisher = {American Institute of Physics Inc.},
   title = {Polarization-induced electrical conductivity in ultra-wide band gap {AlGaN} alloys},
   volume = {109},
   url = {https://pubs.aip.org/apl/article/109/22/222101/31942/Polarization-induced-electrical-conductivity-in},
   year = {2016},
   doi = {10.1063/1.4969062}
}

@article{Tsai1999Theoretical,
    author = {Tsai, Chin-Yi and Chen, Chih-Hsiung and Sung, Tien-Li and Tsai, Chin-Yao and Rorison, Judy M.},
    title = {Theoretical modeling of nonequilibrium optical phonons and electron energy relaxation in GaN},
    journal = {Journal of Applied Physics},
    volume = {85},
    number = {3},
    pages = {1475-1480},
    year = {1999},
    month = {02},
    issn = {0021-8979},
    doi = {10.1063/1.369277},
    url = {https://doi.org/10.1063/1.369277}
}

@article{Fonoberov2003Excitonic,
    author = {Fonoberov, Vladimir A. and Balandin, Alexander A.},
    title = {Excitonic properties of strained wurtzite and zinc-blende {GaN/AlxGa1-xN} quantum dots},
    journal = {Journal of Applied Physics},
    volume = {94},
    number = {11},
    pages = {7178-7186},
    year = {2003},
    month = {12},
    issn = {0021-8979},
    doi = {10.1063/1.1623330},
    url = {https://doi.org/10.1063/1.1623330}
}

@article{Jogai2003InfluenceTransistors,
    title = {{Influence of surface states on the two-dimensional electron gas in AlGaN/GaN heterojunction field-effect transistors}},
    year = {2003},
    journal = {Journal of Applied Physics},
    author = {Jogai, B.},
    number = {3},
    month = {2},
    pages = {1631--1635},
    volume = {93},
    url = {https://pubs.aip.org/jap/article/93/3/1631/293059/Influence-of-surface-states-on-the-two-dimensional},
    doi = {10.1063/1.1530729},
    issn = {0021-8979}
}

@article{Jena2002,
   author = {Debdeep Jena and Sten Heikman and Daniel Green and Dario Buttari and Robert Coffie and Huili Xing and Stacia Keller and Steve DenBaars and James S. Speck and Umesh K. Mishra and Ioulia Smorchkova},
   doi = {10.1063/1.1526161},
   issn = {0003-6951},
   issue = {23},
   journal = {Applied Physics Letters},
   month = {12},
   pages = {4395-4397},
   publisher = {AIP Publishing},
   title = {{Realization of wide electron slabs by polarization bulk doping in graded III–V nitride semiconductor alloys}},
   volume = {81},
   url = {https://pubs.aip.org/apl/article/81/23/4395/114833/Realization-of-wide-electron-slabs-by-polarization},
   year = {2002}
}

@article{Armstrong2019,
   author = {Andrew M. Armstrong and Brianna A. Klein and Albert G. Baca and Andrew A. Allerman and Erica A. Douglas and Albert Colon and Vincent M. Abate and Torben R. Fortune},
   doi = {10.1063/1.5058263},
   issn = {0003-6951},
   issue = {5},
   journal = {Applied Physics Letters},
   month = {2},
   pages = {52103},
   publisher = {American Institute of Physics Inc.},
   title = {{AlGaN polarization-doped field effect transistor with compositionally graded channel from Al0.6Ga0.4N to AlN}},
   volume = {114},
   url = {https://pubs.aip.org/apl/article/114/5/052103/236289/AlGaN-polarization-doped-field-effect-transistor},
   year = {2019}
}

@article{bacaAlrichAlGaNBased2020,
  title = {Al-Rich {{AlGaN}} Based Transistors},
  author = {Baca, A.G. and Armstrong, A.M. and Klein, B.A. and Allerman, A.A. and Douglas, E.A. and Kaplar, R.J.},
  year = {2020},
  journal = {Journal of Vacuum Science and Technology A: Vacuum, Surfaces and Films},
  volume = {38},
  number = {2},
  publisher = {{AVS Science and Technology Society}},
  issn = {07342101 (ISSN)},
  doi = {10.1116/1.5129803},
  langid = {english}
}

@article{brunnerStressEvolutionAlxGa1xN2013,
  title = {Stress Evolution during {{Al}}{\textsubscript{x}}{{Ga}}{\textsubscript{1-x}}{{N}}/{{AlN}} Growth on Sapphire},
  author = {Brunner, F. and Mogilatenko, A. and Kueller, V. and Knauer, A. and Weyers, M.},
  year = {2013},
  journal = {Journal of Crystal Growth},
  volume = {376},
  pages = {54--58},
  issn = {00220248 (ISSN)},
  doi = {10.1016/j.jcrysgro.2013.04.023},
  langid = {english}
}

@article{kangawaInfluenceIntrinsicPoint2025,
  title = {Influence of {{Intrinsic Point Defects Incorporated}} from {{Growth Surface}} on {{Atomic Interdiffusion}} and {{Unintentional Compositional Gradient}} in {{AlGaN}}/{{AlN Heterointerfaces}}},
  author = {Kangawa, Yoshihiro and Kusaba, Akira and Kawamura, Takahiro and Kempisty, Pawel and Ishisone, Kana and Boero, Mauro},
  year = {2025},
  month = feb,
  journal = {Crystal Growth \& Design},
  volume = {25},
  number = {3},
  pages = {740--746},
  issn = {1528-7483, 1528-7505},
  doi = {10.1021/acs.cgd.4c01542},
  urldate = {2025-12-08},
  copyright = {https://creativecommons.org/licenses/by/4.0/},
  langid = {english}
}

@article{liAlNHeteroepitaxySapphire2013,
  title = {{{AlN}} Heteroepitaxy on Sapphire by Metalorganic Vapour Phase Epitaxy Using Low Temperature Nucleation Layers},
  author = {Li, H. and Sadler, T. C. and Parbrook, P. J.},
  year = {2013},
  journal = {International Specialist Meeting on Bulk Nitrides and Related Techniques},
  volume = {383},
  pages = {72--78},
  issn = {00220248 (ISSN)},
  doi = {10.1016/j.jcrysgro.2013.08.024},
  langid = {english}
}

@article{liMassTransportAnalysis2011,
  title = {Mass Transport Analysis of a Showerhead {{MOCVD}} Reactor},
  author = {Li, Hui},
  year = {2011},
  month = mar,
  journal = {Journal of Semiconductors},
  volume = {32},
  number = {3},
  pages = {033006},
  issn = {1674-4926},
  doi = {10.1088/1674-4926/32/3/033006},
  urldate = {2025-12-08},
  langid = {english}
}

@article{mondalInterplayCarrierDensity2025,
  title = {Interplay of Carrier Density and Mobility in {{Al-rich}} ({{Al}}, {{Ga}}) {{N-channel HEMTs}}: {{Impact}} on High-Power Device Performance Potential},
  author = {Mondal, Badal and Pampili, Pietro and Mukherjee, Jayjit and Moran, David and Parbrook, Peter James and Schulz, Stefan},
  year = {2025},
  journal = {APL Electronic Devices},
  volume = {1},
  number = {2},
  publisher = {AIP Publishing},
  doi = {10.1063/5.0277051}
}

@article{moramXrayDiffractionIIInitrides2009,
  title = {X-Ray Diffraction of {{III-nitrides}}},
  author = {Moram, M. A. and Vickers, M. E.},
  year = {2009},
  journal = {Reports on Progress in Physics},
  volume = {72},
  number = {3},
  pages = {036502},
  issn = {00344885 (ISSN)},
  doi = {10.1088/0034-4885/72/3/036502},
  langid = {english}
}

@article{robinsonCrystalTruncationRods1986,
  title = {Crystal Truncation Rods and Surface Roughness},
  author = {Robinson, I. K.},
  year = {1986},
  month = mar,
  journal = {Physical Review B},
  volume = {33},
  number = {6},
  pages = {3830--3836},
  issn = {0163-1829},
  doi = {10.1103/PhysRevB.33.3830},
  urldate = {2026-01-28},
  copyright = {http://link.aps.org/licenses/aps-default-license},
  langid = {english}
}

@article{romanovStressRelaxationMismatched2003,
  title = {Stress Relaxation in Mismatched Layers Due to Threading Dislocation Inclination},
  author = {Romanov, A.E. and Speck, J.S.},
  year = {2003},
  journal = {Applied Physics Letters},
  volume = {83},
  number = {13},
  pages = {2569--2571},
  issn = {00036951 (ISSN)},
  doi = {10.1063/1.1613360},
  langid = {english}
}

@article{saitoImpactGrowthTemperature2025,
  title = {Impact of Growth Temperature on Heterostructure Interface Steepness in Ultraviolet-{{B AlGaN-based}} Laser Diodes},
  author = {Saito, T. and Miyake, R. and Yamada, R. and Imoto, Y. and Maruyama, S. and Sasaki, Y. and Karino, S. and Iwayama, S. and Miyake, H. and Naniwae, K. and Kamiyama, S. and Takeuchi, T. and Iwaya, M.},
  year = {2025},
  journal = {Applied Physics Letters},
  volume = {126},
  number = {8},
  publisher = {American Institute of Physics},
  issn = {00036951 (ISSN)},
  doi = {10.1063/5.0242536},
  langid = {english}
}

@article{singhSurfaceMorphologyEvolution2024,
  title = {Surface {{Morphology Evolution}} of {{AlGaN Microhoneycomb Structures}} during {{Epitaxial Overgrowth}}},
  author = {Singh, S.M. and Zubialevich, V.Z. and Parbrook, P.J.},
  year = {2024},
  journal = {Physica Status Solidi (B) Basic Research},
  volume = {261},
  number = {3},
  publisher = {{John Wiley and Sons Inc}},
  issn = {03701972 (ISSN)},
  doi = {10.1002/pssb.202300471},
  langid = {english}
}

@article{tsaoUltrawideBandgapSemiconductorsResearch2018,
  title = {Ultrawide-{{Bandgap Semiconductors}}: {{Research Opportunities}} and {{Challenges}}},
  author = {Tsao, J.Y. and Chowdhury, S. and Hollis, M.A. and Jena, D. and Johnson, N.M. and Jones, K.A. and Kaplar, R.J. and Rajan, S. and {Van de Walle}, C.G. and Bellotti, E. and Chua, C.L. and Collazo, R. and Coltrin, M.E. and Cooper, J.A. and Evans, K.R. and Graham, S. and Grotjohn, T.A. and Heller, E.R. and Higashiwaki, M. and Islam, M.S. and Juodawlkis, P.W. and Khan, M.A. and Koehler, A.D. and Leach, J.H. and Mishra, U.K. and Nemanich, R.J. and {Pilawa-Podgurski}, R.C.N. and Shealy, J.B. and Sitar, Z. and Tadjer, M.J. and Witulski, A.F. and Wraback, M. and Simmons, J.A.},
  year = {2018},
  journal = {Advanced Electronic Materials},
  volume = {4},
  number = {1},
  publisher = {Blackwell Publishing Ltd},
  issn = {2199160X (ISSN)},
  doi = {10.1002/aelm.201600501},
  langid = {english}
}

@article{wooWideUltrawidebandgapSemiconductors2024,
  title = {From Wide to Ultrawide-Bandgap Semiconductors for High Power and High Frequency Electronic Devices},
  author = {Woo, K. and Bian, Z. and Noshin, M. and Perez Martinez, R. and Malakoutian, M. and Shankar, B. and Chowdhury, S.},
  year = {2024},
  journal = {JPhys Materials},
  volume = {7},
  number = {2},
  publisher = {Institute of Physics},
  issn = {25157639 (ISSN)},
  doi = {10.1088/2515-7639/ad218b},
  langid = {english}
}

@article{yoshikawaUsingLowtemperatureGrowth2023,
  title = {Using Low-Temperature Growth to Resolve the Composition Pulling Effect of {{UV-C LEDs}}},
  author = {Yoshikawa, A. and Zhang, Z. and Kushimoto, M. and Aoto, K. and Sasaoka, C. and Amano, H.},
  year = {2023},
  journal = {Applied Physics Letters},
  volume = {123},
  number = {22},
  publisher = {American Institute of Physics Inc.},
  issn = {00036951 (ISSN)},
  doi = {10.1063/5.0183320},
  langid = {english}
}

@misc{zotero-16966,
  author       = {Mondal, Badal and Pampili, Pietro},
  title        = {{XRDmapAnalysis/xrdpy: xrdpy-v0.0.5}},
  month        = feb,
  year         = 2026,
  publisher    = {Zenodo},
  version      = {version-0.0.5},
  doi          = {10.5281/zenodo.18498887},
  url          = {https://doi.org/10.5281/zenodo.18498887},
}

@article{Birner2007Nextnano:Simulations,
    title = {{nextnano: General Purpose 3-D Simulations}},
    year = {2007},
    journal = {IEEE Transactions on Electron Devices},
    author = {Birner, S. and Zibold, T. and Andlauer, T. and Kubis, T. and Sabathil, M. and Trellakis, A. and Vogl, P.},
    number = {9},
    month = {9},
    pages = {2137--2142},
    volume = {54},
    url = {http://ieeexplore.ieee.org/document/4294186/},
    doi = {10.1109/TED.2007.902871},
    issn = {0018-9383}
}

@article{Trellakis2006TheResults,
    title = {{The 3D nanometer device project nextnano: Concepts, methods, results}},
    year = {2006},
    journal = {Journal of Computational Electronics},
    author = {Trellakis, Alex and Zibold, Tobias and Andlauer, Till and Birner, Stefan and Smith, R. Kent and Morschl, Richard and Vogl, Peter},
    number = {4},
    month = {12},
    pages = {285--289},
    volume = {5},
    publisher = {Springer},
    url = {https://link.springer.com/10.1007/s10825-006-0005-x},
    doi = {10.1007/s10825-006-0005-x},
    issn = {1569-8025}
}
\end{document}

\title{Supplementary Information \\ \vspace*{0.3cm} Barrier--channel intermixing and 2-dimensional electron gas degradation in Al-rich Al(Ga)N/AlGaN high electron mobility transistor heterostructures}
% Force line breaks with \\

\author{Pietro Pampili}%
\email{pietro.pampili@tyndall.ie}
\author{Vitaly Zubialevich}
\author{Badal Mondal}
\affiliation{Tyndall National Institute, University College Cork, Cork T12 R5CP, Ireland}%

\author{Jayjit Mukherjee}
\affiliation{James Watt School of Engineering, University of Glasgow, Glasgow G12 8LT, UK}%

\author{Stefan Schulz}
\affiliation{Tyndall National Institute, University College Cork, Cork T12 R5CP, Ireland}%
\affiliation{School of Physics, University College Cork, Cork T12 YN60, Ireland}%

\author{David Moran}
\affiliation{James Watt School of Engineering, University of Glasgow, Glasgow G12 8LT, UK}%

\author{Peter J. Parbrook}%
\affiliation{Tyndall National Institute, University College Cork, Cork T12 R5CP, Ireland}%
\affiliation{School of Engineering, University College Cork, Western Road, Cork, Ireland}

\date{\today}%
\maketitle
\onecolumngrid

\tableofcontents

%%%%%%%%%%%%%%%%%%%%%%%%%%%%%%%%%%%%%%%%%%%%%%%%%%%%%%%%%%%%%%%%%%%%%%%%%%%%%%
\section{\label{sec:secS1}Nextnano++ simulation details}
One-dimensional (1D) Schr{\"o}dinger-Poisson simulations were performed using the Nextnano++ software (v-1.22.18, RHEL compilation).\cite{Birner2007Nextnano:Simulations,Trellakis2006TheResults} The material parameters used in the simulations are listed in Table~\ref{tab:tableS1}. In the simulations, we assumed a temperature of T = 300 K to match the room-temperature experimental measurements (contactless measurements in the main text) of carrier properties.  

A single-band effective mass model was used for both electron and hole eigenstate calculations. This model is adequate for the present study, as our primary focus is on the properties of the two-dimensional electron gas (2DEG) formed at the barrier-channel interface. We note that the single-band effective mass approximation may introduce minor inaccuracies in the estimation of the two-dimensional hole gas (2DHG) at the substrate-channel interface. However, the relatively long channel thickness of 300 nm spatially separates the 2DHG from the 2DEG interface, thereby minimizing any indirect influence on the 2DEG. Additionally, in wide-bandgap materials such as AlN and GaN, the coupling between conduction- and valence-band states (s$-$p coupling) is weak; consequently, its impact on the accuracy of 2DEG-related results is negligible.\cite{Jogai2003InfluenceTransistors} %Overall, the single-band effective mass model is expected to have a minimal effect on the quantities of interest, which justifies its use in this study. Although a full $8 \times 8$ \textbf{k}$\cdot$\textbf{p} model could yield more refined results, it would significantly increase the computational cost and was therefore not adopted.

The Schr{\"o}dinger quantum calculations were restricted to two quantum regions (QRs), each 40 nm wide, positioned around the substrate-channel and channel-barrier interfaces. The QR width was optimized to ensure that the 2DEG wave‑function tails decay sufficiently before reaching the QR boundaries, while keeping the QR domain as small as possible for computational efficiency. The number of sub-bands included in the quantum calculations was set to 50 for electrons and 100 for holes. These values were optimized to ensure that the high-energy states have vanishing occupations of the respective carriers (electrons and holes). 

A 10 nm-wide Schottky contact was placed on the barrier, and Neumann boundary conditions were applied at the bottom (substrate side) of the simulated structures (see Fig.~\ref{fig:figS1} below and Fig.~6(b) in the main text).\cite{Interplay_Mondal_AED2025} It is important to note that the values of Schottky barrier height (SBH) at the Schottky contact can depend strongly on several factors, including the metal stack (e.g., Ni, Ni/Au, Pt/Au), defects, growth method (e.g, molecular beam epitaxy (MBE) vs. MOVPE), and surface passivation.\cite{Interplay_Mondal_AED2025} However, the detailed nature of these dependencies remains unclear and requires further study. In this work, we adopted the SBH values reported in Ref.~\citenum{Interplay_Mondal_AED2025} (see Table~\ref{tab:tableS1}).

%%%%%%%%%%%%%%%%%%%%%%%%%%%%%%%%%%%%%%%%%%%%%%%%%%%%%%%%%%%%%%%%%%%%%%%%%%%%%%
\section{\label{sec:secS3}Material and model parameters}
Table~\ref{tab:tableS1} summarizes the material parameters used in the Nextnano++ simulations in this work. The parameters for the (Al,Ga)N alloy ($P_{\text{Al}_x \text{Ga}_{1-x} \text{N}}$) were obtained from the corresponding values of the binary compounds GaN and AlN using the quadratic interpolation: 
\begin{equation}
    P_{\text{Al}_x \text{Ga}_{1-x} \text{N}} = x\,P_{\text{AlN}} + (1-x)\,P_{\text{GaN}} - b_{_P}\,x\,(1-x)  \label{eq:eq31}
\end{equation}
where $x$ is the Al mole fraction in Al$_x$Ga$_{1-x}$N, and $b_{_P}$ is the bowing parameter. 
\begin{table}[!ht]
\caption{\label{tab:tableS1}Material parameters for AlN, GaN, and Al$_x$Ga$_{1-x}$N used in this work. Only parameters relevant to our 1D Nextnano++ simulations are listed. All parameters correspond to the unstrained materials. Strain-induced shift in band-edges are accounted for in the simulations through the  deformation potentials. In the table, $\parallel$ denotes the direction parallel to the [0001] c-axis (growth direction, see Fig.~\ref{fig:figS1}), and $\perp$ denotes the direction perpendicular to the [0001] c-axis (in-plane).% \\ \\ Alt text: [Material parameters used in Schrödinger–Poisson simulations for AlN, GaN, and AlGaN alloys. The table summarizes key properties such as bandgap, lattice constants, effective masses, polarization, and elastic constants used to model carrier confinement and band structure.] \\
}

\begin{ruledtabular}
\begin{tabular}{llcccc}
Parameters & Symbol (Unit) & GaN ($x=0$) & AlN ($x=1$)	& Bowing ($b_{_P}$) & Ref. \\
\hline
Schottky barrier height\footnotemark[1] & SBH (eV) & 1.11	& 3.40 & 0 & \citenum{Armstrong2016Polarization-inducedAlloys,Arehart2006EffectCharacteristics} \\
Lattice constants & $a_0$ (\r{A}) at 300 K & 3.189 & 3.112 & 0 & \citenum{Vurgaftman2003BandSemiconductors} \\
 & $c_0$ (\r{A}) at 300 K & 5.185 & 4.982 & 0 & \citenum{Vurgaftman2003BandSemiconductors} \\
Thermal expansion coefficient\footnotemark[2] & $\alpha_{a_0}$ ($10^{-5}$ \r{A} K$^{-1}$) & 1.783 & 1.291 & 0 & \cite{NSMSemiconductors} \\
 & $\alpha_{c_0}$ ($10^{-5}$ \r{A} K$^{-1}$) & 1.644 & 2.626 & 0 & \citenum{NSMSemiconductors} \\
Bandgap energy & E$_\text{g}$ (eV) at 0 K & 3.51 & 6.25	& 0.7 & \citenum{Vurgaftman2003BandSemiconductors} \\ 
Varshni's parameter & $\alpha$ (meV/K)\footnotemark[3] & 0.909 & 1.799 & 0 & \citenum{Vurgaftman2003BandSemiconductors} \\
 & $\beta$ (K) & 830 & 1462 & 0 & \citenum{Vurgaftman2003BandSemiconductors} \\
Crystal-field splitting energy & $\Delta_{cr}$ (eV)  & 0.010 & -0.169 & 0 & \citenum{Vurgaftman2003BandSemiconductors} \\ 
Spin-orbit splitting energy & $\Delta_{so}$ (eV) & 0.017 & 0.019 & 0 & \citenum{Vurgaftman2003BandSemiconductors} \\ 
Average valence band energy\footnotemark[4]$^,$\footnotemark[5]   & $E_{v,av}$ (eV) & $-0.726$ & $-1.526$ & 0 & \citenum{Wei1998CalculatedOrbitals}\\
Electron effective mass & $m_{\parallel} \,(m_0)$ & 0.209 & 0.329 & 0 & \citenum{Rinke2008Consistent}\\
 & $m_{\perp} \,(m_0)$ & 0.186 & 0.322 & 0 & \citenum{Rinke2008Consistent} \\
 Hole effective mass & $m_{\parallel, HH} \,(m_0)$ & 1.1 & 3.53 & 0 & \citenum{NSMSemiconductors} \\
 & $m_{\perp,HH} \,(m_0)$ & 1.6 & 10.42 & 0 & \citenum{NSMSemiconductors}\\
 & $m_{\parallel, LH} \,(m_0)$ & 1.1 & 3.53 & 0 & \citenum{NSMSemiconductors} \\
 & $m_{\perp,LH} \,(m_0)$ & 0.15 & 0.24 & 0 & \citenum{NSMSemiconductors}\\
 & $m_{\parallel, SO} \,(m_0)$ & 0.15 & 0.25 & 0 & \citenum{NSMSemiconductors} \\
 & $m_{\perp,SO} \,(m_0)$ & 1.1 & 3.81 & 0 & \citenum{NSMSemiconductors}\\
Static dielectric constant & $\varepsilon_{s,\parallel} \,(\varepsilon_0)$ & 10.10 & 8.57 & 0 & \citenum{Tsai1999Theoretical,Fonoberov2003Excitonic} \\
Spontaneous polarization constant & $P_{sp}$ (C m$^{-2}$) & $-0.034$ & $-0.090$ & $-0.021$ & \citenum{Vurgaftman2003BandSemiconductors} \\
Elastic constants & $C_{13}$ (GPa) & 106 & 108 & 0 & \citenum{Vurgaftman2003BandSemiconductors} \\
 & $C_{33}$ (GPa) & 398 & 373 & 0 & \citenum{Vurgaftman2003BandSemiconductors} \\
Piezoelectric constants & $e_{31}$ (C m$^{-2}$) & $-0.35$ & $-0.50$ & 0 & \citenum{Vurgaftman2001BandAlloys} \\
 & $e_{33}$ (C m$^{-2}$) & 1.27 & 1.79 & 0 & \citenum{Vurgaftman2001BandAlloys} \\
Deformation potentials\footnotemark[6]$^,$\footnotemark[7] & $a_1$ (eV) & $-4.9$ & $-3.4$ & 0 & \citenum{Vurgaftman2003BandSemiconductors}\\
& $a_2$ (eV) & $-11.3$ & $-11.8$ & 0 & \citenum{Vurgaftman2003BandSemiconductors}\\
& $D_1$ (eV) & $-3.7$ & $-17.1$ & 0 & \citenum{Vurgaftman2003BandSemiconductors}\\
& $D_2$ (eV) & $4.5$ & $7.9$ & 0 & \citenum{Vurgaftman2003BandSemiconductors}\\
& $D_3$ (eV) & $8.2$ & $8.8$ & 0 & \citenum{Vurgaftman2003BandSemiconductors}\\
& $D_4$ (eV) & $-4.1$ & $-3.9$ & 0 & \citenum{Vurgaftman2003BandSemiconductors}
\end{tabular}
\end{ruledtabular}
\footnotetext[1]{Experimental studies have shown a linear relationship between Schottky barrier height and Al content in (Al,Ga)N within the small strain regime.\cite{Qiao2000DependenceFraction,Hahn2013AlNProposal} }

\footnotetext[2]{In Nextnano++, the thermal expansion of the lattice parameter $a$(T) at temperature T is given by: $a (\text{T}) = a (\text{300 K}) + \alpha_{exp} \times  (\text{T}-300)$; where $\alpha_{exp}$ is the linear thermal expansion coefficient. The definition of $\alpha_{exp}$ used here differs slightly from the conventional definitions found in the literature. The relationship between the two can be expressed as: $\alpha_{exp} = a$(300 K) $\times \alpha_{Literature}$; where $\alpha_{Literature}$ is the conventional linear thermal expansion coefficient, expressed in K$^{-1}$.}

\footnotetext[3]{In Nextnano++, the temperature correction to bandgap (E$_{\text{g}}$) is included using Varshni's formula: $\text{E}_{\text{g}} (\text{T}) = \text{E}_{\text{g}} (0) + \delta \text{E}_{\text{g}} (\text{T}) $; where the temperature correction $\delta$E$_{\text{g}}$ (T) for an alloy is interpolated as: 
$ \text{E}_{\text{g, Al}_x\text{Ga}_{1-x}\text{N}} (\text{T},\,x) = x \,\frac{-\alpha_{\text{AlN}}\,\text{T}^2}{\text{T} + \beta_{\text{AlN}}} + (1-x)\,\frac{-\alpha_{\text{GaN}}\,\text{T}^2}{\text{T} + \beta_{\text{GaN}}} - x\,(1-x)\,\frac{-\alpha_{\text{(Al,Ga)N}}\,\text{T}^2}{\text{T} + \beta_{\text{(Al,Ga)N}}} $; 
where $\alpha_{\text{AlN}}$, $\alpha_{\text{GaN}}$, $\beta_{\text{AlN}}$, and $\beta_{\text{GaN}}$ are the Varshni parameters of binary compounds and  $\alpha_{\text{(Al,Ga)N}}$, $\beta_{\text{(Al,Ga)N}}$ are the related alloy bowing. Note that the temperature variation only impacts the conduction bands, since the valence bands act as reference energies for the band offsets.}
\footnotetext[4]{$E_{v,av} = \frac{E_{_{HH}} + E_{_{LH}}  + E_{_{CH}}}{3} - \frac{2}{3}\Delta_{cr}$. The valence band energies for heavy hole (HH), light hole (LH), and crystal-field split-hole (CH) are calculated from $E_{v,av}$.}
\footnotetext[5]{Valence band offset (VBO) between two materials is determined as: $VBO_{v,av} = E_{v,av, \text{material1}} - E_{v,av, \text{material2}}$. }
\footnotetext[6]{Valence and conduction band deformation potentials are used to account for the (biaxial) strain-induced energy shifts in Nextnano++ simulations.}
\footnotetext[7]{$a_1$ and $a_2$ values here are the inter-band deformation potentials, which describe the strain-induced shift of the bandgaps. In Nextnano++, the corresponding valence-band deformation potentials are added to obtain the absolute conduction-band deformation potentials, i.e., $D_\parallel = a_1 + D_1$ and $D_\perp = a_2+D_2$.}
\end{table}

%%%%%%%%%%%%%%%%%%%%%%%%%%%%%%%%%%%%%%%%%%%%%%%%%%%%%%%%%%%%%%%%%%%%%%%%%%%%%%
\section{\label{sec:secS2}Simulated structures details}
\begin{figure}[!h]
    \centering
    \includegraphics[width=\textwidth]{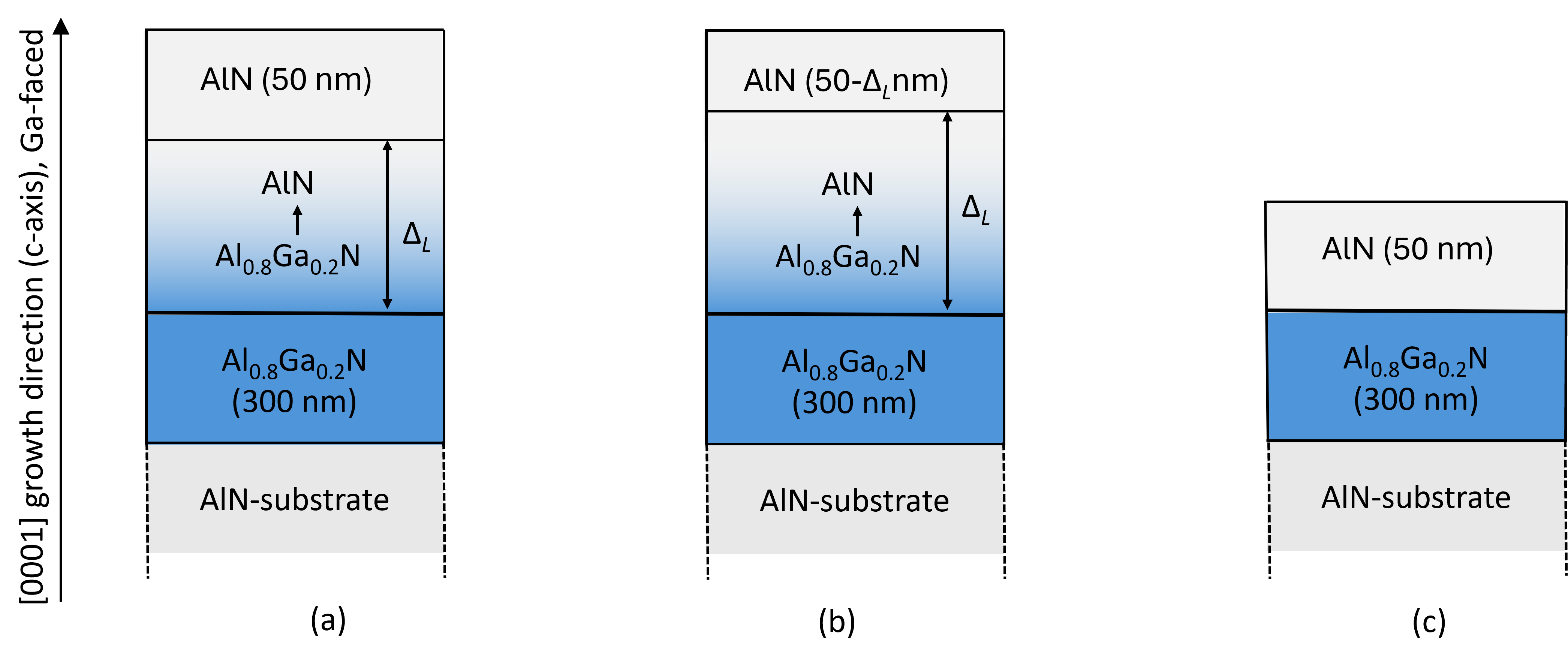}
    \caption{Simulated heterostructures. The Al composition is graded linearly over the length, $\Delta_L$. AlN$\leftarrow$Al$_{0.8}$Ga$_{0.2}$N denotes linear grading from Al$_{0.8}$Ga$_{0.2}$N to AlN. (a) Grading placed between the 50 nm AlN barrier and the (Al,Ga)N channel: AlN(50 nm)/AlN$\leftarrow$Al$_{0.8}$Ga$_{0.2}$N($\Delta_L$ nm)/Al$_{0.8}$Ga$_{0.2}$N(300 nm) systems. (b) Grading applied within the barrier: AlN(50$-\Delta_L$ nm)/AlN$\leftarrow$Al$_{0.8}$Ga$_{0.2}$N($\Delta_L$ nm)/Al$_{0.8}$Ga$_{0.2}$N(300 nm) systems.  (c) Structure with a sharp AlN/(Al,Ga)N barrier-channel interface ($\Delta_L = 0$). %\\ \\ Alt text: [Schematic diagrams of simulated AlGaN/AlGaN heterostructures. The diagrams compare structures with composition grading at the barrier–channel interface, grading within the barrier, and an ideal sharp interface, illustrating how grading length modifies the layer structure.]
    }
    \label{fig:figS1}
\end{figure}

Fig.\ref{fig:figS1} shows the (Al,Ga)N/(Al,Ga)N/AlN-substrate structures simulated in this study (refer to Fig.~6 in the main text). All layers were assumed to be undoped (no intentional or unintentional doping), and a conventional cation-polar [0001] growth direction was considered. Furthermore, all layers were assumed to be pseudomorphically grown on the AlN substrate.  Accordingly, the effect of homogeneous pseudomorphic strain in the channel and barrier layers relative to the AlN substrate was included in the simulations (strain shifts the band edges through deformation potentials).

Two classes of structures were simulated to closely reflect the composition-graded interfaces observed in our experiments (as discussed in the main text): (a) grading placed \textit{between} the AlN barrier and the (Al,Ga)N channel [Fig.\ref{fig:figS1}(a)], and (b) grading applied \textit{within} the AlN barrier [Fig.\ref{fig:figS1}(b)]. The Al composition is \textit{graded linearly} over a distance $\Delta_L$, consistent with the nearly linear profiles measured experimentally (Fig.~4 in the main text).

%\pagebreak
%\FloatBarrier  %PP
%===============================================================================
\bibliography{supplement}% Produces the bibliography via BibTeX.